\documentclass[preprint,12pt]{elsarticle}

\usepackage{graphicx}
\usepackage{pstricks}
\usepackage{color}

\usepackage[figuresright]{rotating}

\newcommand{\lsim}{\smallerthansquiggle}
\def\smallerthansquiggle{\raise.3ex\hbox{$<$\kern-.75em\lower1ex\hbox{$\sim$}}}

\def\greaterthansquiggle{\raise.3ex\hbox{$>$\kern-.75em\lower1ex\hbox{$\sim$}}}
\def\drbar{\ifmmode{\overline{\rm DR}} \else{$\overline{\rm DR}$} \fi}

\newcommand{\AmS}{{\protect\the\textfont2
  A\kern-.1667em\lower.5ex\hbox{M}\kern-.125emS}}

\title{HFOLD - a program package for calculating two-body MSSM Higgs decays at full one-loop level}

\address[HEPHY]{Institute of High Energy Physics, Austrian Academy of Sciences, A-1050 Vienna, Austria}%

\author[HEPHY]{W. Frisch\corref{cor1}}
\ead{frisch@hephy.oeaw.ac.at}
\author[HEPHY]{H. Eberl}
\author[HEPHY]{H. Hluch\'{a}}

\cortext[cor1]{Corresponding author}

\newcounter{bla}

\journal{Computer Physics Communications}

\begin{document}

\begin{frontmatter}

\begin{abstract}
HFOLD (Higgs Full One Loop Decays) is a Fortran program package for calculating all MSSM Higgs two-body decay widths and the corresponding branching ratios at full one-loop level. 
The package is done in the SUSY Parameter Analysis convention and supports the  
SUSY Les~Houches Accord input and output format.
\vspace{1pc}
\end{abstract}


\end{frontmatter}
{\bf PROGRAM SUMMARY}

\begin{small}
\noindent
{\em Manuscript Title:} HFOLD - a program package for calculating two-body MSSM Higgs decays at full one-loop level\\
{\em Authors:} Wolfgang Frisch, Helmut Eberl, Hana Hlucha                                                \\
{\em Program Title:} HFOLD                                   \\
{\em Journal Reference:}                                      \\
{\em Catalogue identifier:}                                   \\
{\em Licensing provisions:} none                              \\
{\em Programming language:} Fortran 77                        \\
{\em Computer:} Workstation, PC                               \\
{\em Operating system:} Linux                                 \\
{\em RAM:} 524288000                                         \\ 
{\em Number of processors used:}                              \\
{\em Supplementary material:}                                 
{\em Classification:} 11.1                                     \\
{\em External routines/libraries:} SLHALib 2.2                \\
{\em Subprograms used:} LoopTools 2.2                         \\
{\em Nature of problem:}\\
A future high--energy $e^{+}e^{-}$ linear collider will be the best
environment for the precise measurements of masses, cross sections, branching ratios, etc..  
Experimental accuracies are expected at the per-cent down to the per-mille
level.  These must be matched from the theoretical
side. Therefore higher order calculations are mandatory. 
\\
{\em Solution method:}\\
This program package calculates all MSSM Higgs two-body decay widths and the corresponding branching ratios at full one-loop level.
The renormalization is done in the \drbar scheme following the SUSY Parameter Analysis convention. The program supports the  
SUSY Les~Houches Accord input and output format.
\\
{\em Restrictions:}\\
   \\
{\em Unusual features:}\\
   \\
{\em Additional comments:}\\
   \\
{\em Running time:} \\

\end{small}

\section{Introduction}

The Minimal Supersymmetric Standard Model (MSSM) is the most extensively studied extension 
of the Standard Model (SM) of elementary particles. Supersymmetry (SUSY) 
provides a solution to
the so called hierarchy problem and furthermore, in the context of this work, it is a
renormalizable
theory. If the MSSM is realized in nature, supersymmetric particles will be produced at the LHC. 
However, even if SUSY is discovered, it will still be a long way to determine the parameters
of the underlying model, which would shed light on the mechanism of SUSY breaking. A future high--energy $e^{+}e^{-}$ linear collider will be the best
environment for the precise measurements of masses, cross sections, branching ratios, etc..  
Experimental accuracies are expected at the per-cent down to the per-mille
level \cite{R2,R3A,R2B}.  These must be matched from the theoretical
side. Therefore higher order calculations are mandatory. 

For the decays of the MSSM Higgs bosons, 
the one-loop corrections due to gluon and gluino exchange (SQCD) are known analytically,
see e.g. \cite{djouadi,djouadi2,abartl,abartl2,coarasa}.
Full one--loop calculations were done e.g. in
\cite{hollik,eberl1,eberl2,eberl3,eberl4,eberl5, eberl6, dabelstein, heinemeyer, heinemeyer2}. 
For calculating the full (including electroweak corrections) 
one-loop  decay widths 
automatic tools for generating all Feynman graphs, and subsequently the squared matrix elements, are strongly needed. 

There are a few program packages available for the automatic computation of amplitudes at full one--loop level in the MSSM: FeynArts/FormCalc \cite{feynarts-formcalc}, SloopS \cite{SLOOPS_Higgs,SLOOPS_Sf} and GRACE/SUSY-loop \cite{GRACE/SUSY-loop}. SloopS and GRACE/SUSY-loop also perform renormalization at one--loop level. However, so far there is no publicly available code for the two--body Higgs decays at full one--loop level in the MSSM.  
Therefore, we have developed the Fortran code HFOLD \cite{SUSYFOLD}. It follows the renormalization prescription of the SUSY Parameter Analysis project (SPA) \cite{SPA} and supports the SUSY Les Houches Accord (SLHA) input and output format \cite{SLHA1}.
The package HFOLD (Higgs Full One-Loop Decays) computes all 
two-body decay widths and the corresponding branching ratios of the three neutral and charged Higgs bosons
at full one-loop level. 

This paper is organized in the following way: First we shortly recapitulate the Higgs sector in the 
MSSM. Then we will discuss the renormalization used in the program. We will compare the total and partial decay widths of the Higgs bosons at 
the SPS1a' point with existing programs. The last section will be the program manual.


\section{MSSM Higgs sector at tree-level}

\subsection{Masses and mixing angles}

In the MSSM two chiral Higgs superfields with opposite hypercharge are necessary
to keep the theory anomaly free. Two Higgs doublets are also necessary in order to
give separately masses to down-type fermions and up-type quarks.

The scalar components of the two complex isospin Higgs doublets
\begin{displaymath}
H_1 =  \left( \begin{array}{cc} H_1^0\\ H_1^- \end{array}\right)\,  , \quad
H_2 =  \left( \begin{array}{cc} H_2^+\\ H_2^0 \end{array}\right)\,  ,  \nonumber
\end{displaymath}
represent eight scalar degrees of freedom (d.o.f.) and have hypercharges $Y(H_{1,2}) = \mp 1$.
After spontaneous electroweak symmetry breaking, their neutral components receive
vacuum expectation values, $\langle H_1^0 \rangle = v_1$ and $\langle H_2^0 \rangle = v_2$.
The absolute value $v^2 = v_{1}^{2} + v_{2}^{2}$ can be determined from the measurements of
e.g.  $m_W$ and the SU(2) coupling $g$, but
$\tan\beta = {v_2 \over v_1}$ remains a free parameter.
There remain five physical Higgs bosons, two neutral CP even ones,
$h^0$ and $H^0$ and one neutral CP odd field $A^0$ and two charged Higgs bosons $H^\pm$.
The physical states $h^0$ and $H^0$ are mixtures described by the mixing angle $\alpha$.
The remaining three d.o.f. are 'eaten' by the longitudinal components of the now massive vector 
bosons $Z^0$  and $W^\pm$. 

At tree-level only two free parameters describe the Higgs sector. In the MSSM
usually the parameters $m_{A^0}$ and $\tan\beta$
are chosen.
The other three Higgs boson masses and 
the mixing angle $\alpha$ can be expressed at tree--level by $m_Z$ and $m_W$,
e.g.  $m^2_{H^+} = m_W^2 +  m^2_{A^0}$. 
Contrary to the SM, the Higgs self-interactions are completely fixed by EW parameters.
At tree-level the mass of the lightest Higgs boson $h^0$ cannot be larger than $m_Z$. 
This value is already ruled out by LEP2. Fortunately, radiative corrections push the theoretical limit
up to $m_{h^0} \lsim 135$~GeV with the leading contributions from top and stop loops proportional to
$m_t^4/m_W^2$.\\
 
\newpage
\subsection{Decay patterns and some properties}

\begin{figure}
{\setlength{\unitlength}{1mm}
\begin{center}
\mbox{
\begin{picture}(75,60)(0,0)
\put(3,0){\includegraphics[width=7cm]{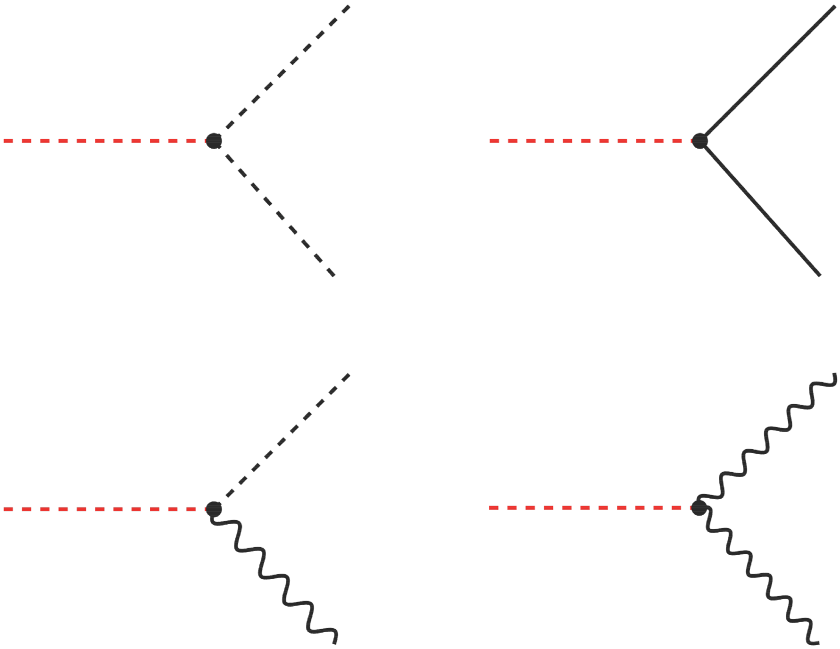}}

\put(3,53.25){\makebox(0,0)[bl]{\small a)}}
\put(3,43.5){\makebox(0,0)[bl]{\small scalar}}
\put(28,53){\makebox(0,0)[c]{\rotatebox{45}{\small scalar}}}
\put(30,36){\makebox(0,0)[c]{\rotatebox{312}{\small scalar}}}

\put(44,53.5){\makebox(0,0)[bl]{\small b)}}
\put(44,43.5){\makebox(0,0)[bl]{\small scalar}}
\put(67.5,51.5){\makebox(0,0)[c]{\rotatebox{45}{\small fermion}}}
\put(69.5,36.5){\makebox(0,0)[c]{\rotatebox{312}{\small fermion}}}

\put(3,23){\makebox(0,0)[bl]{\small c)}}
\put(3,13){\makebox(0,0)[bl]{\small scalar}}
\put(28,22){\makebox(0,0)[c]{\rotatebox{45}{\small scalar}}}
\put(30,5){\makebox(0,0)[c]{\rotatebox{312}{\small vector}}}

\put(44,23){\makebox(0,0)[bl]{\small d)}}
\put(44,13){\makebox(0,0)[bl]{\small scalar}}
\put(68,22){\makebox(0,0)[c]{\rotatebox{45}{\small vector}}}
\put(70,5){\makebox(0,0)[c]{\rotatebox{312}{\small vector}}}

\end{picture}
}
\end{center}
}
\vspace*{-5mm}
\caption{Four possibilities of two-body decays of a scalar particle}
\label{decaystructures}
\end{figure}
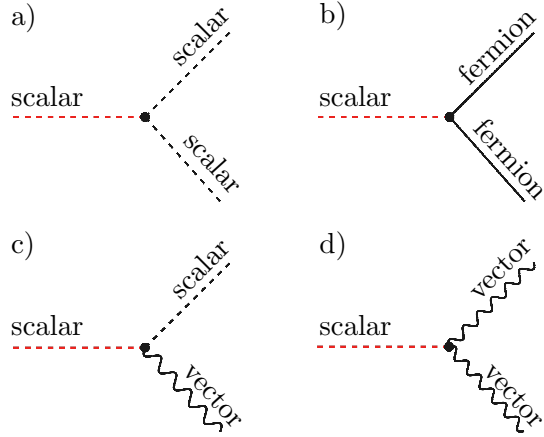

As fermion number is conserved we only have four possibilities
of  Feynman graphs (at any loop level) for a two-body decay
of a scalar: the decay into two scalars, into two fermions, into a scalar and 
a vector particle, and into two vector particles, see Fig.~\ref{decaystructures}.
In the case of Higgs bosons the following decays are calculated:  
\begin{eqnarray*}
{\rm  Fig.~\ref{decaystructures}a:}\quad
\phi &\to&  \tilde f_i \, \tilde f_{j}^{*}\,,\\
 H^+  &\to&  \tilde f_i \, \tilde f_{j}^{'*}\,,\\
 H^0 &\to& h^0 \, h^0 \, , \,  A^0 \, A^0\,, \\
{\rm  Fig.~\ref{decaystructures}b:}\quad
\phi &\to& f \, \bar{f}\,,\\
\phi &\to& {\tilde \chi}^0_k \, {\tilde \chi}^0_l\,(k, l = 1,\ldots,4)\,,\\ 
\phi &\to& {\tilde \chi}^+_r \, {\tilde \chi}^-_s\,(i,j,r,s = 1,2)\,, \\ 
 H^+ &\to& f \, \bar{f}' \,,\\
 H^+ &\to& {\tilde \chi}^0_k \, {\tilde \chi}^+_s\,, \\
%
{\rm  Fig.~\ref{decaystructures}c:}\quad
 A^0 &\to& h^0 \, Z^0\,,H^0 \, Z^0 \,,\\
 H^+ &\to& h^0  \, W^+\, , \,  H^0 \, W^+\,,\\
{\rm  Fig.~\ref{decaystructures}d:}\quad
 H^0 &\to& Z^0 \, Z^0 , \, \,  W^+ \, W^-\,,\\
\phi &\to&  \gamma \gamma,\, g g,\,\gamma Z^0 \quad {\rm (loop~induced)}\,,
\end{eqnarray*}
$\phi =  h^0,\, H^0,\, A^0$ and $f\!=\!\nu_l,\;e,\;\mu,\;\tau,\;u,\;d,\;c,\;s,\;t,\;b$, $f^{'}$ denotes the isospin partner to f, e.g. 
$f=t,\;f^{'}=b$, $\tilde{f}$ and $\tilde{f}^{'}$ denote the SUSY partners of $f$ and $f^{'}$, ${\tilde \chi}^0$ and ${\tilde \chi}^\pm$ are the neutralinos and charginos, respectively.
The Higgs bosons couple to fermions via their Yukawa couplings. 
Therefore, the branching ratio (BR) into top quark(s) 
is large, if the decay is kinematically allowed. The BRs of $h^0 \to \bar b\,b$ and to $\tau^+ \tau^-$ are dominant, 
especially for large $\tan\beta$. The decays into the third generation sfermions may become dominant when they 
are kinematically possible. The decays into quarks and squarks can have large 
one-loop SQCD corrections. The decays into charginos and/or neutralinos can have
significant one-loop contributions from the third generation
(s)fermions depending on the gaugino/higgsino mixing.

Decoupling limit:
In case of $m_{A^0} \gg m_{Z^0}$ the masses of $H^0$, $A^0$, and $H^+$ become degenerate, 
\begin{displaymath}
m_{h^0} \ll m_{H^0} \sim m_{A^0} \sim m_{H^+}\, .\nonumber
\end{displaymath}
This limit is already reached to a good approximation for $m_{A^0} \sim 300$~GeV.
Furthermore, the  $(h^0, H^0)$ mixing angle can be expressed by $\alpha \to \beta - \pi/2$. Thus, the properties
of the lightest Higgs boson $h^0$ are almost indistinguishable from those of the SM Higgs boson. 
As a consequence, the couplings to the heavier Higgs bosons vanish at 
tree-level, e.g.  the $H^+ W^- h^0$ coupling is $\propto \cos(\beta - \alpha) \to 0$.

\section{Calculation at full one-loop level}
The definition of the MSSM parameters is not unique beyond the leading
order and depends on the renormalization scheme. Therefore, a
well-defined theoretical framework was proposed
within the so-called SPA (SUSY Parameter Analysis) project
\cite{SPA}. The "SPA convention" provides a clear base for
calculating masses, mixing angles, decay widths and production
processes. It also provides a clear definition of the fundamental
parameters using the \drbar (dimensional reduction)
renormalization scheme. These fundamental parameters can be extracted from
future collider data. The formulae for the wave function and mass counterterms (CTs) for sfermions, 
fermions and vector bosons in the on-shell scheme derived
from their renormalization conditions can be found 
e.g. in \cite{denner,chiprode+e-,sfermionprode+e-}. 

The code of HFOLD is derived in the SPA convention 
in the general linear $R_\xi$ gauge for the $W^{\pm}$ and $Z^0$-boson. 
All amplitudes are generated by using the tool 
FeynArts (FA) and the Fortran code is produced with the help
of FormCalc (FC). For that purpose we imported all necessary formulae for the CTs into a FA model file.

\begin{figure}[h]
\includegraphics[width=12.1cm]{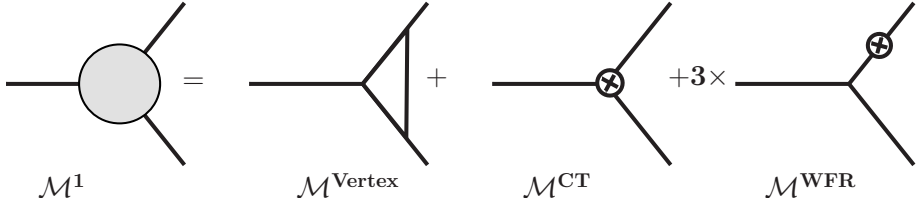}
\vspace*{5mm}
\caption{One-loop renormalization procedure of a 1 to 2 process schematically}
\label{3point-ren}
\label{fig:renproc}
\end{figure}

The renormalized one-loop amplitude is the sum of the 
tree-level amplitude and the one-loop contributions, see Fig~\ref{3point-ren}.

The tree-level couplings are given at the scale $Q$, implying that there are no coupling CTs. The  \drbar scheme is defined by setting the UV divergence parameter $\Delta =0$.
We however work with $\Delta \ne 0$ and take for the coupling CTs
only the parts $\propto \Delta$. In case the renormalized amplitudes are finite it is a proof for RGE invariance of the ordinary \drbar scheme.

The vertex corrections and all 
selfenergy contributions except the diagonal wave function corrections 
can be directly calculated with FA/FC.

Since there are many decay channels it was worthwhile to develop an automatic code generator
at Mathematica level. First of all, it was necessary to 
work out all counterterms (in Mathematica form) for the whole MSSM. 
The idea is, not to have all MSSM couplings (which are 
more than 300 ones) at one-loop level hard coded  in the MSSM model file
of FA, but to calculate locally the amplitudes with the wave function and the coupling CTs
(see Fig.~\ref{fig:renproc}). 

For each external particle we get a contribution to the 
wave function CTs amplitude by multiplying the bare fields
with the corresponding wave function renormalization constants.
The amplitude for the coupling CTs is obtained in the following way: 
First we calculate the tree-level amplitude, then we shift all tree-level couplings by their corresponding counterterms
 $\delta g_i$,  $g_i \to g_i + \delta g_i$ and then take into account only terms linear in $\delta g_i$. 

The total two-body Higgs decay width can be written in one-loop approximation as
\begin{eqnarray*}
\Gamma  &=& N_C \times kin \times \left( |{\cal M}_{0}|^2 + 2\;{\rm Re}({\cal M}_{0}^{\dagger} {\cal M}_1)\right)\,,\\
kin &=& \frac{\kappa(m_{0}^{2},m_{1}^{2},m_{2}^{2})}{16 \pi m_{0}^{3}}\,,
\end{eqnarray*}
with the totally symmetric {\textit{K\"allen function}} $\kappa
(x, y, z) = \sqrt{(x-y-z)^2 - 4 y z}$ and the color factor $N_C
= 3$ for decays into quarks and squarks and \mbox{$N_C = 1$} for decays into other particles,
respectively.

$ {\cal M}_1$ denotes the UV finite one-loop amplitude. The prefactor $kin$ is
a function of the on-shell masses of the incoming Higgs boson and outgoing particles only. 
Massless particles in loops can cause so-called infrared (IR) divergences
in  $\Gamma$. For this purpose, a regulator mass $\lambda$ for the photon and gluon is introduced. 
Adding then real photon or gluon radiation cancels these divergences.

\section{Input parameters}
HFOLD is designed to be applied to SUSY models like mSUGRA, GMSB or AMSB, where the 
low energy model parameters are given at some scale Q. The low energy spectrum is derived from a few parameters defined at a high scale using renormalization group equations.
At the program start HFOLD reads the spectrum, where the Yukawa couplings, the gauge couplings $g_1, g_2, g_3$, the soft breaking terms,
the VEVs, $m_{A^{0}}, \tan \beta$, $\mu$ and the on-shell Higgs masses are taken as input parameters.
The input parameters are understood as running parameters in the \drbar scheme at the scale Q.
In loops we are free to use \drbar masses because the difference is of higher order in perturbation theory. Since our renomalization is done in the \drbar scheme 
the coupling counterterms contain only UV-divergent parts. Therefore we do not fix $\delta m_W$ with $G_{Fermi}$ as input parameter.
In the Higgs sector we use $m_{A^{0}}$ and the running $\tan \beta$ as inputs. We can then simply derive the \drbar running Higgs mixing angle $\alpha$ at the scale Q. 
We do not take $\alpha_{eff}$ as input parameter because we consider our calculation a self--consistent one--loop expansion.

\section{Resummation of $\tan \beta$}
The down-type fermions couple to the up-type Higgs doublet with radiative corrections by
\begin{equation}
-y_b H_d^0 \bar b b - y_b \Delta_b \cot\beta H_u^0 \bar b b\;.
\end{equation}  
The selfenergy $\Delta_b$ is  proportional to $\tan \beta$ and can be enhanced for large values of $\tan \beta$.
This term can be resummed (in the effective potential approach) by replacing the bottom Yukawa coupling \cite{carenaDeltab} with
\begin{equation}
y_b \to \frac{y_b}{1 + \Delta_b}\;.
\label{resumyuk}
\end{equation}
The resummation can also be 
performed in the diagrammatic approach \cite{nierste}.
Different renormalization schemes correspond to different choices of counterterms. Therefore 
the analytic form of the $\tan \beta$ enhanced corrections depend on the chosen renormalization scheme. 
In the on-shell scheme one takes the measured bottom mass as input parameter. 
The choice of $\delta m_b$ fixes $\delta y_b$ by
\begin{equation}
y_b=\frac{m_b}{v_d} \rightarrow \delta y_b=\frac{\delta m_b}{v_d}\,.
\end{equation}
The quark mass counterterm $\delta m_b$ is a source of $\tan \beta$-enhanced corrections. 
The selfenergy $\Sigma^{RL}(m_b)$ contains terms proportional to $y_b \sin \beta$ and is therefore $\tan \beta$ enhanced,
\begin{eqnarray}
\Sigma^{RL}=m_b \Delta_b\,,\\
\Delta_b=\Delta_{b}^{\tilde{g}}+\Delta_{b}^{\tilde{\chi}^{\pm}}+\Delta_{b}^{\tilde{\chi}^{0}}\,.
\end{eqnarray}
In leading order this means : $\delta m_b = -\Sigma_{b}^{RL}=-m_b \epsilon_b \tan \beta$.
We write the bare Yukawa couplings as $y_b^{(0)}= y_b + \delta y_b $,
where $y_b$ is the renormalized coupling and $ \delta y_b $ is the
counterterm.  The choice of $ \delta m_b$ fixes $ \delta y_b $ through
\begin{equation}
  \delta y_b=\frac{\delta m_b}{v_d}=-y_b\epsilon_b\tan\beta .
  \label{YukCT}
\end{equation}
The supersymmetric loop effects encoded in $\epsilon_b$ enter physical
observables only through $\delta y_b$. Choosing e.g.\ a minimal subtraction scheme like the \drbar
scheme for $\delta m_b$ removes the $\tan\beta$-enhanced terms 
and there is nothing to resum anymore. 
Since we do not use the measured bottom mass as input, the resummation of $\tan \beta$ is absent in our approach.
However, the resummation eq.~(\ref{resumyuk}) is implemented in the code and can be turned on.

\subsection{Gauge used}

\noindent The gauge fixing Lagrangian in the general linear $R_\xi$ gauge is given by
\begin{displaymath}
{\cal L}^{GF} = -{1 \over \xi_W} F^+ F^- {1 \over \xi_A} |F^A|^2\,, \quad A = Z,
\gamma, g\, ,
\end{displaymath}
with $F^+ = \partial_\mu W^{\mu +} + i \xi_W m_W G^+$, $F^Z = \partial_\mu Z^\mu + \xi_Z m_Z G^0,
F^\gamma = \gamma_\mu A^\mu$, and $F^g = \gamma_\mu G^{a \mu}$.\\

\noindent The Higgs-ghost propagators are $i/(q^2 - \xi_V m_V^2)$ and 
the vector-boson propagator reads
\begin{displaymath}
D^{\mu \nu}_V=\frac{-i\left(g^{\mu \nu} -(1-\xi_V)\frac{q^{\mu} q^{\nu}}{q^{2}-\xi m_V^{2}}
\right)}{q^{2}- m_V^{2}}\, .
\end{displaymath}
The ${\xi}$-dependent part is a product of two propagators leading to 
a (n+1)-point loop integral. Performing a decomposition into partial fractions, 
it can be split into a form with single propagators only, 
\begin{displaymath}
 D^{\mu \nu}_V = 
 \frac{-i \, g^{\mu \nu}}{q^{2} \!-\! m_V^{2}} +
 \frac{i}{m_V^{2}}\!
\left(\!\frac{q^{\mu}q^{\nu}}{q^{2} \!-\! m_V^{2}}  -  \frac{q^{\mu}q^{\nu}}{q^{2} \!-\! \xi m_V^{2}} \!\right)\,.
\end{displaymath}
We have implemented this second form into FA in order to check the gauge independence
for $W$ and $Z$. For the massless particles $\gamma$ and gluon we get derivatives of loop
integrals. In these cases it is possible to proof gauge invariance analytically.

\subsection{Photon/Gluon radiation}
The IR divergences can be removed using soft bremsstrahlung or by adding the 
corresponding 1 to 3 process with a massless particle (hard brems\-strahlung).
Soft radiation is
proportional to the tree-level width but dependent on the energy cut $\Delta E$
of the radiated--off particle. It is automatically included in FC, see the formulae in \cite{denner} . 
For an 1 to 3 process with a massless particle 
the three-body phase space can still be integrated out analytically.
We have implemented this radiation by using self-derived generic formulae for all four
graphs in  Fig.~\ref{decaystructures} where every charged line
can radiate off a photon (or a gluon for colored particles). 
The IR convergent total width is then given by
\begin{displaymath}
\Gamma^{\rm total} = \Gamma(\phi_0 \to p_1 p_2) + \Gamma(\phi_0 \to p_1 p_2 \, \gamma/g)\, .
\nonumber
\end{displaymath}
For the simplest case, the decay into two scalars, $ \Gamma(\phi \to \phi_1 \phi_2 \, \gamma/g)$ is proportional to
\begin{eqnarray*}
    &\int& \overline{|\mathcal{M}|^2}\to  \\
      & -& 4 |g_{\rm tree}|^2\, \times\,  \big[ \hphantom{+} g_0^2 m_0^2 I_{{0 0}} + g_0 g_1 (m_0^2 + m_1^2 - m_2^2) I_{{1 0}} + g_1^2 m_1^2 I_{11} \\
      &+&  g_0 g_2 (m_0^2 -m_1^2 + m_2^2) I_{20} + g_2^2 m_2^2 I_{22} + g_1 g_2 (m_0^2 - m_1^2 - m_2^2) I_{21}  \\
      &+&  g_0 (g_1 + g_2) I_0 + g_1 (g_0 - g_2) I_1 + g_2 (g_0 - g_1)I_2 \big]\,.
     \nonumber
\end{eqnarray*}
The 'bremsstrahlung integrals' $I$ are given in  \cite{denner}.
The integrals $I_{ij}$ depend on $\log\lambda$, here $\lambda$ is the auxiliary mass for $\gamma/g$.
For the cases scalar $\to$ fermion + fermion with one fermion mass zero 
(e.g. $H^{+} \to  \tau^{+} \nu_{\tau}$), we have derived special formulae for the bremsstrahlung integrals.
The other formulae can be found explicitly in the program code in the file bremsstrahlung.F.

\section{Program manual}

\subsection{Requirements}
\begin{itemize}
\setlength{\itemsep}{-1mm}
 \item {Fortran 77  (g77, ifort77)}
 \item{C compiler (e.g. gcc)}
 \item{LoopTools~\cite{looptools}}
\end{itemize}

\subsection{About version 1.0}
\begin{itemize}
\setlength{\itemsep}{-1mm}
 \item{The CKM matrix is set diagonal}
 \item{Real SUSY input parameters}
\end{itemize}

\subsection{Installation}

\begin{enumerate}
\item Download the file \textbf{hfold.tar} at 
\begin{quote}
http://www.hephy.at/tools
\end{quote}
\item expand the file, go to the folder hfold/SLHALib-2.2 and type
\begin{quote}
\tt{./configure}
\end{quote}
\begin{quote}
\tt{make}
\end{quote}
\item to create the Fortran code for {\tt hfold}, go back to the folder hfold and type
\begin{quote}
\texttt{./configure}
\end{quote}
\begin{quote}
\tt{make}
\end{quote}
\item To run HFOLD type
\begin{quote}
\tt{hfold}
\end{quote}
\end{enumerate}

\subsection{The input file {\tt hfold.in}}
\begin{enumerate}
\setlength{\itemsep}{-1mm}
 \item \textbf{name of the spectrum (SLHA format)}
\\
 \item \textbf{Higgs boson = 1,2,3,4,5}\\$1=h^0,\;2=H^0,\;3=A^0,\;4=H^{+},\;5=All$
\\
 \item \textbf{contribution = 0,1,2}\\
0 = tree--level calculation \\
1 = full one--loop calculation\\
2 = SQCD (only diagrams with a gluon/gluino are taken into account) 
\\
 \item \textbf{bremsstrahlung = 0,1,2}\\0 = off, 1 = hard bremsstrahlung, 2 = soft bremsstrahlung 
\\ 
\item \textbf{resummation of bottom yukawa coupling = 0,1}\\0 = off, 1 = on
\\
 \item \textbf{esoftmax}\\cut on the soft photon(gluon) energy, if soft strahlung is used
\\
 \item \textbf{name of output-file}
\end{enumerate}

\section{Comparison HFOLD with HDECAY 3.53 and FEYNHIGGS 2.7.4}

\noindent
{\bf SPS1a' point:}\\
\noindent
In the following we show some results for the mSUGRA point proposed in the SPA project \cite{SPA},
$(M_{1/2}, M_0, A_0) = (250, 70, -300)$~GeV, ${\rm sign}(\mu) = +1$, and $\tan\beta = 10$. Our comparison
with other programs is based on the same input file with the MSSM spectrum given in 
SUSY Les Houches accord from\cite{SLHA1} created by {\tt SPheno3.0beta}. A list of available decay programs 
is given at {\tt http://home.fnal.gov/~skands/slha/}. 
In the following tables the Higgs bosons partial and total decay widths are compared to HDECAY 3.53 and FeynHiggs2.7.4.
In FeynHiggs 2.7.4 the decays into fermions are at full one-loop level. HDECAY 3.53 \cite{hdecay} has implemented higher order QCD and some EW corrections. Most of these corrections are mapped into running masses. 
\begin{center} 
\begin{table}[htbp]
\begin{tabular}{cccccc}
 $\boldmath{h^{0}}$ & \textbf{HF-tree}& \textbf{HF-SQCD}& \textbf{HF-full}& \textbf{FH 2.7.4}& \textbf{HD 3.53}  \\ 
$\boldmath{\Gamma^{total}}$ & 1.9&3.0& 2.8& 3.2&3.7\\ 
 \end{tabular}
\caption{Comparison of the total decay widths of the CP-even Higgs boson $h^0$ (in MeV)}
\end{table}
%
\begin{table}[htbp]
\begin{tabular}{cccccc}
 $\boldmath{H^{0}}$ & \textbf{HF-tree}& \textbf{HF-SQCD}& \textbf{HF-full}& \textbf{FH 2.7.4}& \textbf{HD 3.53}  \\ 
$\boldmath{\Gamma^{total}}$ & 0.8389&1.0171&1.0274&0.9890&1.0495\\ 
 \end{tabular}
 \caption{Comparison of the total decay widths of the CP-even Higgs boson $H^0$}
\end{table}
%
\begin{table}[htbp]
\begin{tabular}{cccccc}
$\boldmath{A^{0}}$ & \textbf{HF-tree}& \textbf{HF-SQCD}& \textbf{HF-full}& \textbf{FH 2.7.4}& \textbf{HD 3.53}  \\ 
$\boldmath{\Gamma^{total}}$ & 1.2471&1.4405&1.5256&1.4183&1.4139\\ 
 \end{tabular}
 \caption{Comparison of the total decay widths of the CP-odd Higgs boson $A^0$}
\end{table}

%
 \begin{table}[htbp]
\begin{tabular}{cccccc}
$\boldmath{H^{+}}$ & \textbf{HF-tree}& \textbf{HF-SQCD}& \textbf{HF-full}& \textbf{FH 2.7.4}& \textbf{HDECAY}  \\ 
$\boldmath{\Gamma^{total}}$ & 0.7534&0.9057&0.8948&0.7875&0.9671\\ 
\end{tabular}
\caption{Comparison of the total decay widths of the charged Higgs boson $H^+$} 
\end{table}

\begin{table}[htbp]
\begin{tabular}{cccccccccc}
\boldmath{$h^{0}$} & $\textbf{BR-tree}$ & \textbf{HF-tree}& \textbf{HF-SQCD}& \textbf{HF-full}& \textbf{FH 2.7.4}& \textbf{HD 3.53}  \\ 
\boldmath{$b \bar{b}$} &0.8044&0.0015&0.0026&0.0024&0.0025&0.0029  \\ 
\boldmath{$\tau \bar{\tau}$} &0.1544&0.0003&0.0003&0.0003&0.0003&0.0003  \\ 
\boldmath{$c \bar{c}$} &0.0403&0.0001&0.0001&0.0001&0.0001&0.0001  \\ 
\end{tabular}
\label{table:h0decaywidths}
\caption{Comparison of the partial decay widths of the CP-even Higgs boson $h^0$}
\end{table}

\begin{table}[htbp]
\begin{tabular}{cccccccccc}
 \boldmath{$H^{0}$} & $\textbf{BR-tree}$ & \textbf{HF-tree}& \textbf{HF-SQCD}& \textbf{HF-full}& \textbf{FH 2.7.4}& \textbf{HD 3.53}  \\ 
\boldmath{$b \bar{b}$} &0.5546&0.4652&0.6262&0.6216&0.6283&0.6466  \\ 
\boldmath{$\tau \bar{\tau}$} &0.1058&0.0887&0.0887&0.0914&0.0983&0.0909  \\ 
\boldmath{$t \bar{t}$} &0.0549&0.0460&0.0631&0.0564&0.0607&0.0937  \\ 
\boldmath{$\tilde{\chi}^{0}_{1} \tilde{\chi}^{0}_{2}$} &0.0539&0.0452&0.0452&0.0465&0.0429&0.0442  \\ 
\boldmath{$\tilde{\chi}^{+}_{1} \tilde{\chi}^{-}_{1}$} &0.0515&0.0432&0.0432&0.0527&0.0528&0.0568  \\ 
\boldmath{$\tilde{\tau}_{1} \tilde{\tau}_{1}$} &0.0212&0.0177&0.0177&0.0184&0.0183&0.0095  \\ 
\boldmath{$\tilde{\tau}_{1} \tilde{\tau}_{2}$} &0.0206&0.0173&0.0173&0.0191&0.0183&0.0262  \\ 
\boldmath{$\tilde{\chi}^{0}_{2} \tilde{\chi}^{0}_{2}$} &0.0205&0.0172&0.0172&0.0206&0.0210&0.0225  \\ 
\boldmath{$\tilde{\chi}^{0}_{1} \tilde{\chi}^{0}_{1}$} &0.0172&0.0144&0.0144&0.0140&0.0122&0.0127  \\ 
\end{tabular}
\label{table:HHdecaywidths}
\caption{Comparison of the partial decay widths of the CP-even Higgs boson $H^0$}
\end{table}

\begin{table}[htbp]
\begin{tabular}{cccccccccc}
 \boldmath{$A^{0}$} & $\textbf{BR-tree}$ & \textbf{HF-tree}& \textbf{HF-SQCD}& \textbf{HF-full}& \textbf{FH 2.7.4}& \textbf{HD 3.53}  \\ 
\boldmath{$b \bar{b}$} &0.3741&0.4665&0.6282&0.6250&0.6269&0.6439  \\ 
\boldmath{$\tilde{\chi}^{+}_{1} \tilde{\chi}^{-}_{1}$} &0.1800&0.2245&0.2245&0.2862&0.2395&0.2389  \\ 
\boldmath{$t \bar{t}$} &0.0862&0.1074&0.1389&0.1289&0.1881&0.1815  \\ 
\boldmath{$\tilde{\chi}^{0}_{1} \tilde{\chi}^{0}_{2}$} &0.0755&0.0942&0.0942&0.0972&0.0890&0.0871  \\ 
\boldmath{$\tilde{\chi}^{0}_{2} \tilde{\chi}^{0}_{2}$} &0.0729&0.0909&0.0909&0.1166&0.0975&0.0955  \\ 
\boldmath{$\tau \bar{\tau}$} &0.0713&0.0889&0.0889&0.0919&0.0980&0.0911  \\ 
\boldmath{$\tilde{\tau}_{1} \tilde{\tau}_{2}$} &0.0225&0.0280&0.0280&0.0297&0.0292&0.0272  \\ 
\boldmath{$\tilde{\chi}^{0}_{1} \tilde{\chi}^{0}_{1}$} &0.0170&0.0212&0.0212&0.0205&0.0181&0.0183  \\ 
\end{tabular}
\label{table:A0decaywidths}
\caption{Partial decay widths of $A^0$}
\end{table}

\begin{table}[htbp]
\begin{tabular}{cccccccccc}
\boldmath{$H^{+}$} & $\textbf{BR-tree}$ & \textbf{HF-tree}& \textbf{HF-sqcd}& \textbf{HF-full}& \textbf{FH 2.7.4}& \textbf{HDECAY 3.53}  \\ 
\boldmath{$t \bar{b}$} &0.6171&0.4649&0.6170&0.5989&0.5060&0.6850  \\ 
\boldmath{$\tilde{\chi}^{+}_{1} \tilde{\chi}^{0}_{1}$} &0.1712&0.1290&0.1290&0.1306&0.1194&0.1228  \\ 
\boldmath{$ \tau \nu_{\tau}$} &0.1203&0.0906&0.0906&0.0944&0.0922&0.0927  \\ 
\boldmath{$\tilde{\nu_\tau} \tilde{\tau}_1$} &0.0809&0.0610&0.0610&0.0643&0.0630&0.0581  \\ 
\end{tabular}
\label{table:Hpdecaywidths}
\caption{Comparison of the decay widths of the charged Higgs boson $H^+$}
\end{table}

\end{center}

\clearpage
The screen output when running HFOLD is as following:
\begin{verbatim}
 
                         _    _ ______ ____  _      _____    
                        | |  | |  ____/ __ \| |    |  __ \   
                        | |__| | |__ | |  | | |    | |  | |  
                        |  __  |  __|| |  | | |    | |  | |  
                        | |  | | |   | |__| | |____| |__| |  
                        |_|  |_|_|    \____/|______|_____/   1.0
 
                     Higgs Full One Loop Decays by W. Frisch,
                     H. Eberl, H. Hlucha
 
 
 error 0
 abort 173248520
 nslhadata 5504
 ====================================================
   FF 2.0, a package to evaluate one-loop integrals
 written by G. J. van Oldenborgh, NIKHEF-H, Amsterdam
 ====================================================
 for the algorithms used see preprint NIKHEF-H 89/17,
 'New Algorithms for One-loop Integrals', by G.J. van
 Oldenborgh and J.A.M. Vermaseren, published in 
 Zeitschrift fuer Physik C46(1990)425.
 ====================================================
 ffxdb0: IR divergent B0', using cutoff   1.
 ffxc0i: infra-red divergent threepoint function,
 working with a cutoff   1.
 Flags:
 ----------------------------
 Susyqcd calculation
 ----------------------------
 resummation of bottom yukawa coupling off
 ----------------------------
 Using onshell Higgs masses from : SPS1aprime.spc                          
 ----------------------------
 hard bremsstrahlung on
 ----------------------------
 the output SLHA will be written to : output.slha                             
 ----------------------------
 ============================
      Decay Table :
      Total width :  0.905677243
      H+ -> mu+ nu_mu        :   0.320441E-003 / BR :     0.00
      H+ -> tau+ nu_tau      :   0.906196E-001 / BR :     0.10
      H+ -> c sb             :   0.349004E-003 / BR :     0.00
      H+ -> t bb             :   0.617035E+000 / BR :     0.68
      H+ -> ~chi_10 ~chi_1+  :   0.128997E+000 / BR :     0.14
      H+ -> ~chi_30 ~chi_1+  :   0.699102E-003 / BR :     0.00
      H+ -> h W+             :   0.277014E-002 / BR :     0.00
      H+ -> ~nu_eL ~el_2+    :   0.126903E-002 / BR :     0.00
      H+ -> ~nu_muL ~mu_1+   :   0.237453E-003 / BR :     0.00
      H+ -> ~nu_muL ~mu_2+   :   0.124702E-002 / BR :     0.00
      H+ -> ~nu_tauL ~tau_1+ :   0.609862E-001 / BR :     0.07
      H+ -> ~nu_tauL ~tau_2+ :   0.114467E-002 / BR :     0.00

\end{verbatim}
   
\section{Acknowledgments}

The authors acknowledge support from EU under the MRTN-CT-2006-035505 network program and from the
"Fonds zur F\"orderung der wissenschaft\-lichen Forschung" of Austria, project No. P 18959-N16 and project No. I 297-N16.
We thank Walter Majerotto for helpful comments.

\bibliographystyle{elsarticle-num}

\end{document}